\documentclass[prl, twocolumn, showpacs, superscriptaddress]{revtex4-2}
\usepackage{amsmath, amssymb}
\usepackage{graphicx}
\renewcommand{\section}[1]{{\par\it #1.---}\ignorespaces}

\begin{document}
\title{Identifying the ground state phases by spin-patterns in the Shastry-Sutherland model}
\author{Yun-Tong Yang}
\affiliation{School of Physical Science and Technology $\&$ Lanzhou Center for Theoretical Physics, Lanzhou University, Lanzhou 730000, China}
\affiliation{Key Laboratory of Quantum Theory and Applications of MoE $\&$ Key Laboratory of Theoretical Physics of Gansu Province, Lanzhou University, Lanzhou 730000, China}
\author{Fu-Zhou Chen}
\affiliation{School of Physical Science and Technology $\&$ Lanzhou Center for Theoretical Physics, Lanzhou University, Lanzhou 730000, China}
\affiliation{Key Laboratory of Quantum Theory and Applications of MoE $\&$ Key Laboratory of Theoretical Physics of Gansu Province, Lanzhou University, Lanzhou 730000, China}
\author{Hong-Gang Luo}
\email{luohg@lzu.edu.cn}
\affiliation{School of Physical Science and Technology $\&$ Lanzhou Center for Theoretical Physics, Lanzhou University, Lanzhou 730000, China}
\affiliation{Key Laboratory of Quantum Theory and Applications of MoE $\&$ Key Laboratory of Theoretical Physics of Gansu Province, Lanzhou University, Lanzhou 730000, China}

\begin{abstract}
Exploring the influence of frustration on the phases and related phase transitions in condensed matter physics is of fundamental importance in uncovering the role played by frustration. In the two-dimensional square lattice, a minimal frustration has been formulated in 1981 as the Shastry-Sutherland (SS) model described by competitions between the nearest-neighbor bond ($J_1$) and the next-nearest-neighbor one ($J_2$). In the two limits of $\alpha=J_2/J_1$, i.e. $\alpha \ll 1$ and $\alpha \gg 1$, the corresponding phases are the N{\'e}el antiferromagnet (AFM) and the dimer-singlet(DS). Unfortunately, the intermediate regime remains controversial, and the nature of transition from the N{\'e}el AFM to the intermediate state is also unclear. Here we provide a pattern language to explore the SS model and take the lattice size $L=4 \times4$ with periodic boundary condition. We firstly diagonalize the Hamiltonian in an operator space to obtain all fundamental spin-patterns and then analyze their energy and occupancy evolutions with the frustration parameter $\kappa=\alpha / (1+\alpha)$. Our results indicate that the intermediate regime is characterized by diagonal two-domain spin-pattern while the N{\'e}el AFM state has a diagonal single-domain and the DS has mixings of diagonal single- and four-domain. While the transition from the DS to the intermediate phase occurred around $\alpha_c = 1.5$ is the first-order in nature, consistent with that in literature, the one from the intermediate phase to the AFM is clearly seen around $\alpha_c = 1.277$, where it has a reversal of the contributions from the single- and two-domain patterns to the ground state. The result indicates that the pattern language is powerful in identifying the possible phases in frustrated models.
\end{abstract}
\maketitle

\section{Introduction}
Frustrated magnetism has become an extremely active field in condensed matter physics due to finding potentially new states or new properties of this kind of matter \cite{Lacroix2011}. As a minimal frustrated model in the two-dimensional (2D) square lattice, the Shastry-Sutherland (SS) model has been extensively studied. This model was originally proposed in 1981 \cite{Shastry1981} as an exactly solvable antiferromagnetic model with nearest-neighbor ($J_1$) and next-nearest-neighbor ($J_2$) couplings. A dimer-singlet (DS) state forms an exact eigenstate at all couplings and is the ground state in a region where $J_2$ dominates \cite{Shastry1981, Miyahara1999}. Interests in this model mainly come from two aspects. The first is the discovery of the quasi-2D material $SrCu_2(BO_3)_2$ \cite{Kageyama1999, Miyahara1999} in 1999 because the spin network of $SrCu_2(BO_3)_2$ is topologically equivalent to the SS model \cite{Miyahara1999}. The second is that $SrCu_2(BO_3)_2$ is a typical spin-gap system \cite{Kageyama1999} which may have relevance to the pseudogap behavior observed in the high-$Tc$ cuprates superconductivity \cite{Bednorz1986}. 

The SS model is in the N{\'e}el antiferromagnetic (AFM) state and the DS state at both ends of phase diagram. However, the intermediate regime with maximal frustration remains controversial. Different candidate ground states were proposed by different approaches. For examples, a direct transition from the N{\'e}el AFM phase to the DS phase was obtained by means of series expansion \cite{Zheng1999, Erwin2000} and variational projected entangled pair states algorithm \cite{Isacsson2006}; a helical state was addressed by using Schwinger boson mean-field theory \cite{Albrecht1996} and gauge-theoretic analysis \cite{Chung2001}; a plaquette valence bond state (VBS) was first reported by means of series expansion in 2000 \cite{Koga2000, Takushima2001}, then other methods such as exact diagonalization (ED) \cite{Andreas2002, Nakano2018}, density matrix renormalization group (DMRG) \cite{Moukouri2008, Lee2019} and tensor network method \cite{Corboz2013, Wang2023, Ning2023} also give the same result. Futhermore, the plaquette VBS is further divided into full plaquette VBS \cite{Zayed2017, Boos2019, Cui2023} and empty plaquette VBS \cite{Lee2019, Yang2022, Wang2022}; a columnar-dimer state, an incommensurate spin-density wave phase and a topologically ordered state were also proposed by means of series expansion \cite{Zheng2001}, field theory \cite{Carpentier2001} and ED \cite{Ronquillo2014}, respectively. Very recently, the coexistence of a plaquette VBS and a quantum spin liquid state between Ne{\'e}l AFM and DS was proposed by using DMRG \cite{Yang2022}, ED \cite{Wang2022} and functional renormalization group \cite{Ahmet2022}. Moreover, the nature of phase transition, particularly between the N{\'e}el AFM phase and the intermediate phase, is also unclear. Is it a second-order transition \cite{Koga2000} or weakly first-order transition \cite{Corboz2013}, or other forms such as a deconfined quantum critical point \cite{Lee2019, Ning2023}?

Here we use the pattern language to provide some insights into the above controversies. The pattern language was proposed in our previous works \cite{Yang2022b, Yang2022c, Yang2023a, Yang2023b, Yang2023c, Yang2023d, Yang2023e}. We have applied this language to one-dimensional and 2D frustrated spin models, namely, the axial next-nearest-neighbor Ising model (ANNNI) \cite{Yang2023c} and the spin-1/2 $J_1$-$J_2$ Heisenberg model on a square lattice \cite{Yang2023e}. The results provide an explicit physical picture for the maximal frustration regime of these two models. Naturally, we still employ the pattern language to study the SS model. For simplicity, we take lattice size $L=L_x\times L_y$ ($L_x=L_y=4$) under periodic boundary condition (PBC). Firstly, we diagonalize the Hamiltonian in the operator space consisting of spin components $(S^x_i, -iS^y_i, S^z_i)$ ($i$ is the lattice site). Six kinds of patterns are obtained in which three kinds are important to the ground state and low-lying excited states due to lower energies, namely, the patterns $\lambda_1$-$\lambda_3$ with diagonal single-domain, the patterns $\lambda_4$-$\lambda_{15}$ with diagonal two-domain and the patterns $\lambda_{16}$-$\lambda_{27}$ with mixings of diagonal single- and four-domain. Secondly, we identify the ground state phases by analyzing patterns' evolution. The results indicate that the phase diagram is roughly divided into three regions: the N{\'e}el AFM, the intermediate regime and the DS, which are dominated by the patterns $\lambda_1$-$\lambda_3$, the patterns $\lambda_{4}$-$\lambda_{15}$ and the patterns $\lambda_{16}$-$\lambda_{27}$, respectively. The transition from the DS to the intermediate regime is a first-order phase transition, occurred around $\alpha_c \sim 1.5 $. The transition from the N{\'e}el AFM to the intermediate one is also seen around $\alpha_c \sim 1.277 $ which corresponds the evolution from the patterns $\lambda_1$-$\lambda_3$ to the patterns $\lambda_{4}$-$\lambda_{15}$. In the following we provide the detail to confirm our statements. 

\section{Model and Method}
The Hamiltonian of the SS model reads
\begin{equation}
\hat{H}' = J_1\sum_{\langle i,j\rangle}\boldsymbol{\hat S_i \cdot \hat S_{j}} + J_2\sum_{\langle\langle i,j\rangle\rangle}\boldsymbol{\hat S_i \cdot \hat S_{j}}, \label{ssm0}
\end{equation}
where $S_i$ are $S=1/2$ operators, $\langle i,j\rangle$ represents the nearest-neighbor bonds in the square lattice, and $\langle\langle i,j\rangle\rangle$ represents the next-nearest-neighbor bonds belonging to orthogonal dimer. In what follows, we take $J_1$ as units of energy and introduce $\alpha = J_2/J_1$, representing a dimensionless frustration parameter. Thus Eq. (\ref{ssm0}) is reformulated as $\hat{H}'= \frac{J_1}{2} \hat{H}$, where 
\begin{small}
\begin{equation}
\hat{H} = \sum_{\langle i,j\rangle} \left(\boldsymbol{\hat S_i \cdot \hat S_{j} + \hat S_j \cdot \hat S_{i}}\right) + \alpha \sum_{\langle\langle i,j\rangle\rangle} \left(\boldsymbol{\hat S_i \cdot \hat S_{j} + \hat S_j \cdot \hat S_i}\right). \label{ssm1}
\end{equation}
\end{small}
Furthermore, Eq. \eqref{ssm1} can be written equivalently as a $3L\times 3L$ matrix
\begin{eqnarray}
&&\hat{H} = \left(
\begin{array}{cccccccccc}
\hat S^x_1& i\hat S^y_1& \hat S^z_1& \hat S^x_2& i\hat S^y_2& \hat S^z_2 & \cdots & \hat S^x_{L}& i\hat S^y_{L}& \hat S^z_{L}
\end{array}
\right)\nonumber\\
&&\times
\left(
\begin{array}{cccccccccc}
0 &0 &0 &1 & 0 &0 &\cdots &\alpha &0 &0 \\
0 &0 &0 &0 & 1 &0 &\cdots &0 &\alpha &0 \\
0 &0 &0 &0 &0 &1 &\cdots &0 &0 &\alpha \\
1 &0 &0 &0 &0&0 &\cdots &0 &0 &0 \\
0 &1 &0 &0 &0 &0 &\cdots &0 &0 &0 \\
0 &0 &1 &0 &0 &0 &\cdots &0 &0 &0 \\
\vdots &\vdots &\vdots &\vdots &\vdots &\vdots &\ddots &\vdots &\vdots &\vdots \\
\alpha &0 &0 &0 &0&0&\cdots &0 &0 &0 \\
0 &\alpha &0 &0 &0&0 &\cdots &0 &0 &0 \\
0 &0 &\alpha &0 &0&0 &\cdots &0 &0 &0
\end{array}
\right)\times \label{ssm2}\\
&&\left(
\begin{array}{cccccccccc}
\hat S^x_1& -i\hat S^y_1& \hat S^z_1& \hat S^x_2& -i\hat S^y_2& \hat S^z_2& \cdots & \hat S^x_{L}& -i\hat S^y_{L}& \hat S^z_{L}
\end{array}
\right)^T, \nonumber
\end{eqnarray}
where the superscript $T$ denotes transpose. This matrix can be diagonalized to obtain eigenvalues and corresponding eigenfunctions $\{\lambda_n, u_n\}  (n = 1, 2, \cdots, 3L)$, which define the patterns marked by $\lambda_n$. Thus the SS model Hamiltonian is rewritten as
\begin{equation}
\hat{H} = \sum_{n=1}^{3L} \lambda_n \hat{A}^\dagger_n \hat{A}_n, \label{ssm3a} 
\end{equation}
where each pattern $\lambda_n$ composes of single-body operators
\begin{equation}
\hat{A}_n = \sum_{i=1}^{L}\left[u_{n,3i-2} \hat S^x_{i} + u_{n,3i-1} (-i\hat S^y_{i}) + u_{n,3i} \hat S^z_{i}\right]. \label{ssm3b}
\end{equation}
In order to cover the whole coupling regime, we define another parameter 
\begin{equation}
\kappa=\frac{\alpha}{1+\alpha},
\end{equation}
where $\kappa \in [0,1)$ corresponds to $\alpha \in [0,+\infty)$. In the following we take $\kappa$ as the frustration parameter.

\begin{figure*}[tbp]
\begin{center}
\includegraphics[width =1.9 \columnwidth]{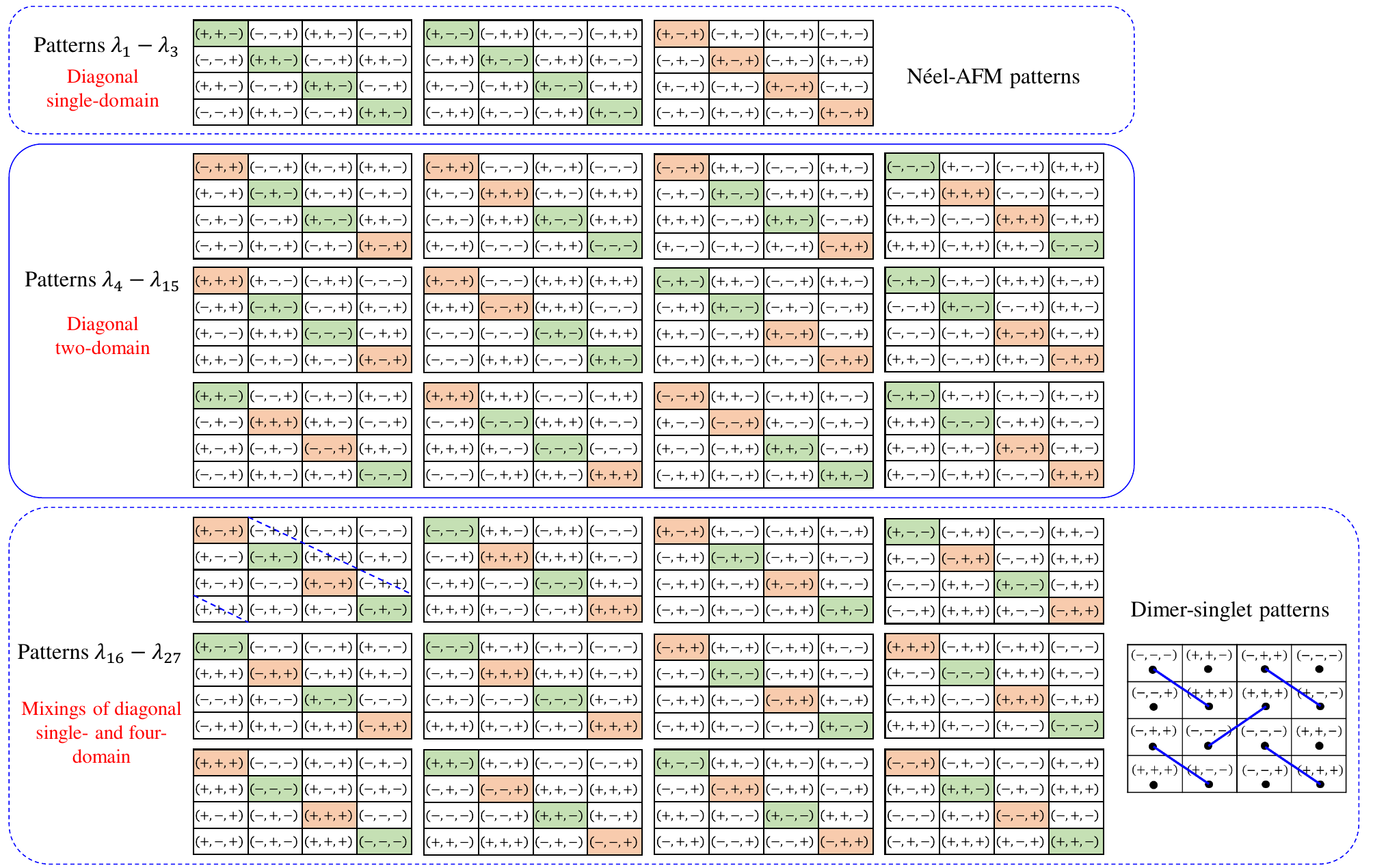}
\caption{The patterns and their relative phases obtained by diagonalization in operator space, marked by the single-body operators Eq. \eqref{ssm3b} with $(\pm,\pm,\pm)$ denoting the signs of $(u_{n,3i-2},u_{n,3i-1},u_{n,3i})$. The four diagonal sites are marked by orange and green representing $(\pm,\pm,+)$ and $(\pm,\pm,-)$. In addition, the patterns are free of a total phase factor $e^{i\pi}$ but their relative phases remain fixed. The bottom right is the orthogonal dimer bonds of the SS lattice.} \label{fig1}
\end{center}
\end{figure*}

\begin{figure}[tbp]
\begin{center}
\includegraphics[width =0.8 \columnwidth]{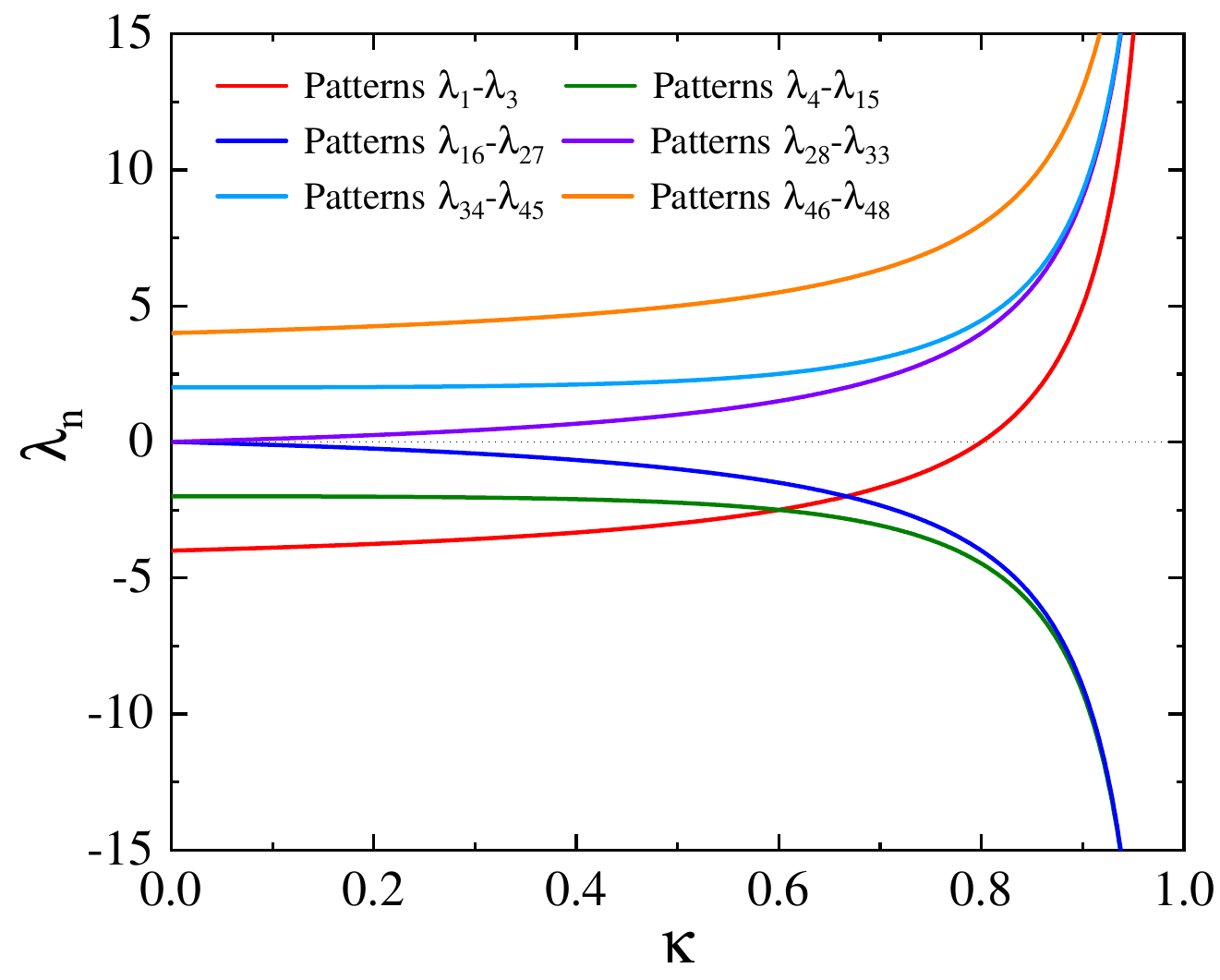}
\caption{The eigenenergies of the patterns as functions of the frustrated parameter $\kappa$ for $L=4 \times 4$ lattice size.}\label{fig2}
\end{center}
\end{figure}

\section{Patterns' Information} 
Firstly, we explore the properties of the patterns marked by relative signs of eigenfunctions $(u_{n,3i-2},u_{n,3i-1},u_{n,3i})$ and their eigenvalues $\lambda_n$ as functions of $\kappa$, shown in Fig. \ref{fig1} and Fig. \ref{fig2}. Six kinds of patterns with different degeneracy are obtained: the first kind of pattern (red line) is threefold-degeneracy which is named the patterns $\lambda_1$-$\lambda_3$; the second (green line) and the third pattern (blue line) are both twelvefold-degeneracy, named the patterns $\lambda_4$-$\lambda_{15}$ and the patterns $\lambda_{16}$-$ \lambda_{27}$, respectively. These three kinds of patterns have negative eigenenergies in the most region of the $\kappa$, thus are important to the ground state and the low-lying excited states. The remainders, i.e. the patterns $\lambda_{28}$-$\lambda_{33}$, the patterns $\lambda_{34}$-$\lambda_{45}$ and $\lambda_{46}$-$\lambda_{48}$, have positive eigenenergies which have zero or less contributions to the cases we consider here. Therefore, we only list the details in Fig. \ref{fig1} for first three kinds of patterns. 

Now we check domain numbers along diagonals of the SS model, as done in the square $J_1$-$J_2$ Heisenberg model \cite{Yang2023e}. For each site $i$, there are three spin components $(S^x_i, -iS^y_i, S^z_i)$ in which $S^z_i$ represents spin states (spin-up or spin-down) and other two components represent dynamical information (spin flip). We label spin states with different colors according to the plus or minus sign of the coefficients $u_{n,3i}$ and choose one diagonal as representative, as shown in Fig. \ref{fig1}. The patterns $\lambda_1$-$\lambda_3$ have a feature of diagonal single-domain and spins show opposite alignment. Therefore the patterns $\lambda_1$-$\lambda_3$ are identified as the N{\'e}el AFM state. The patterns $\lambda_4$-$\lambda_{15}$ have a feature of diagonal two-domain. The patterns $\lambda_{16}$-$ \lambda_{27}$ not only have diagonal four-domain but also contain diagonal single-domain (dashed blue line). For this pattern, we mark the orthogonal dimer bonds of the SS lattice with blue lines, shown in the lower right of Fig. \ref{fig1}. Obviously, the patterns $\lambda_{16}$-$ \lambda_{27}$ have a feature of DS state.

After identifying the characteristic features of the patterns, it is interesting to check the corresponding eigenenergies of these patterns, shown in Fig. \ref{fig2}. At small $\kappa$, the patterns $\lambda_1$-$\lambda_3$ have lower energies and this regime is the N{\'e}el AFM phase in nature. As $\kappa$ increases, the patterns $\lambda_1$-$\lambda_3$ first compete with the patterns $\lambda_{4}$-$\lambda_{15}$, then compete with the patterns $\lambda_{16}$-$\lambda_{27}$. The two intersections indicate that the system takes place two phase transitions or crossovers. To obtain more detailed information about phase transitions and the intermediate regime, we do further calculations.

\begin{figure}[tbp]
\begin{center}
\includegraphics[width = \columnwidth]{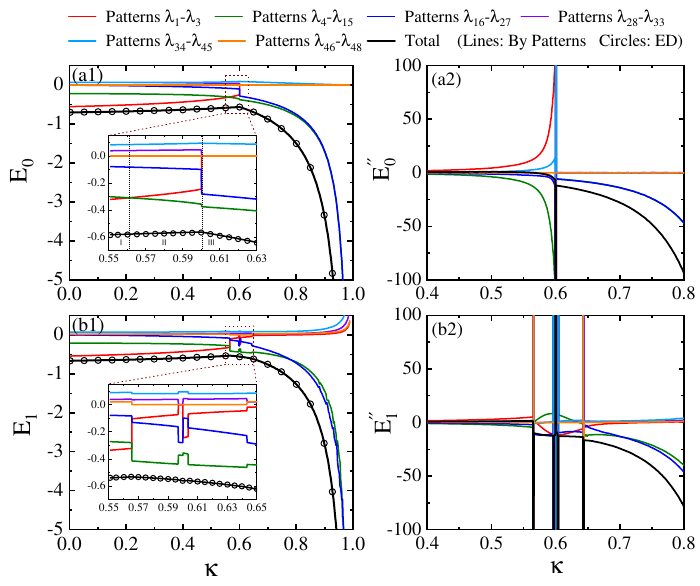}
\caption{(a1) $\&$ (b1) The ground state and the first excited state energies as functions of the frustration parameter $\kappa$ (thick black solid lines) and their pattern dissections (thin color solid lines). ED results (circles) are also presented. The vertical dashed-lines in the inset of (a1) marks two cross points, $\kappa_{c1} \approx 0.5609$ ($\alpha_{c1} \approx 1.277$) and $\kappa_{c2} \approx 0.6$ ($\alpha_{c2} \approx 1.5$), respectively, which divide the frustration parameter $\kappa$ into three regions: I, the N{\'e}el AFM; II, the diagonal two-domain dominant; and III, the DS state. (a2) $\&$ (b2) The second derivatives of the corresponding energy levels (thick black solid lines) and their pattern dissections (thin color solid lines). }\label{fig3}
\end{center}
\end{figure}

We firstly obtains the matrix $\left[\hat{A}_n\right]_{\{S^z_i\},\{S^z_i\}^{\prime}} = \langle\{S^z_i\}|\hat{A}_n|\{S^z_i\}^{\prime}\rangle$ by inserting into the complete basis $|\{S^z_i\}\rangle$ and then Eq. (\ref{ssm3a}) can be solved by diagonalizing the matrix with elements
\begin{eqnarray}
&& \langle \{S^z_i\}|\hat{H}|\{S^z_i\}^{\prime}\rangle = \sum_{n=1}^{3L} \lambda_n \nonumber\\
&& \hspace{1cm}\times \sum_{\{S^z_i\}^{\prime\prime}} \left[\hat{A}^\dagger_n\right]_{\{S^z_i\},\{S^z_i\}^{\prime\prime}}\left[\hat{A}_n\right]_{\{S^z_i\}^{\prime\prime},\{S^z_i\}^{\prime}}.\label{Ising4}
\end{eqnarray}
Figure \ref{fig3} (a1) $\&$ (b1) present the results for the ground state and the first excited state energies as functions of $\kappa$, as shown by thick black solid lines, respectively. Direct numerical ED results are also presented to check the validity of the pattern picture, shown as circles. The exact agreement between them is noticed, which is not surprising since no any approximation has been introduced. 

\section{Identifying the Phases in the SS Model}
\begin{figure}[tbp]
\begin{center}
\includegraphics[width = \columnwidth]{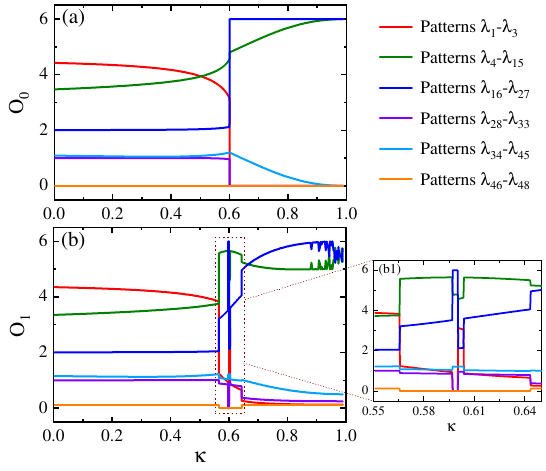}
\caption{(a) The patterns' occupancy in the ground state and (b) the first excited state as functions of $\kappa$. The jitter in (b) is due to the degeneracy of the first excited state with higher excited states.}\label{fig4}
\end{center}
\end{figure}

Fig. \ref{fig3} (a1) is the dissection of the ground state energy which can be roughly divided into three regions according to two intersections, as shown in the inset of (a1). Region I is the N{\'e}el AFM phase, where the patterns $\lambda_1$-$\lambda_3$ dominate over others. The patterns $\lambda_4$-$\lambda_{15}$ also have a remarkable contribution. Other patterns almost have no contribution to this region. Region II is dominated by the patterns $\lambda_{4}$-$\lambda_{15}$. In this region, the contribution from the patterns $\lambda_1$-$\lambda_3$ fades away and that from the patterns $\lambda_4$-$\lambda_{15}$ grows up slowly. Therefore, this region is mainly characterized by the patterns $\lambda_{4}$-$\lambda_{15}$, which have diagonal two-domain. The cross point is about $\kappa_{c1} \approx 0.5609$ ($\alpha_{c1} \approx 1.277$). Region III is the DS phase. The cross point is about $\kappa_{c2} \approx 0.6$ ($\alpha_{c2} \approx 1.5$). In this region, the patterns $\lambda_{4}$-$\lambda_{15}$ and the patterns $\lambda_{16}$-$\lambda_{27}$ are of equal importance, as shown in Fig. \ref{fig3} (a1). However, from the patterns' occupancy defined as $O = \langle\Psi|\hat{A}^\dagger_n \hat{A}_n|\Psi\rangle$ where $|\Psi\rangle$ is the wavefunction, the patterns $\lambda_{16}$-$\lambda_{27}$ have more large occupancy than the patterns $\lambda_{4}$-$\lambda_{15}$, shown in Fig. \ref{fig4} (a). Thus, the physics of this region is dominated by the patterns $\lambda_{16}$-$\lambda_{27}$, identified as the DS state. In addition, the patterns $\lambda_{34}$-$\lambda_{45}$ have a minor positive contribution caused by the interplay between the quantum fluctuation and the magnetic exchange interactions. Other patterns almost have no contributions to the ground state energy.

Fig. \ref{fig3} (a2) is the second derivative of the ground state and their pattern dissections. Obviously, at $\kappa_{c2} \approx 0.6$, a first-order phase transition occurs. This is consistent with the results reported in literature \cite{Albrecht1996, Koga2000, Chung2001, Andreas2002, Corboz2013, Lee2019}. At $\kappa_{c1} \approx 0.5609$, the second derivative is continuous which suggests that the phase transition does not occur at $\kappa_{c1}$, but other ways such as series expansions or correlation functions suggest that the phase transition occurs \cite{Koga2000, Corboz2013, Lee2019, Ning2023}. This is the origin of controversies. The pattern picture clearly shows what happened near $\kappa_{c1}$: the patterns $\lambda_4$-$\lambda_{15}$ defeat the patterns $\lambda_1$-$\lambda_3$ and the model evolves from diagonal single-domain to diagonal two-domain, i.e. from the N{\'e}el AFM to the intermidiate phase. In addition, it is worth mentioning that the two cross points are roughly consistent with those reported in the literature \cite{Darradi2005, Lou2012, Corboz2013, Nakano2018, Lee2019, Nakano2022, Ning2023}, irrespective of the small lattice size we consider here.  

We also analyzed the first excited state by the same way, as shown in Fig. \ref{fig3} (b1) and Fig. \ref{fig4} (b). The contributions of different patterns have similar behaviors, but the first excited state occurs more first-order phase transitions. At $\kappa \approx 0.6$ ($\alpha \approx 1.5$), there occur three first-order phase transitions in quick succession. Besides, at $\kappa \approx 0.5657$ ($\alpha \approx 1.3025$) and $\kappa \approx 0.6432$ ($\alpha \approx 1.8026$), there also take place first-order phase transitions. Figs. \ref{fig3} (b2) provides the second derivative of the first excited state energy, which confirm the above statement. The patterns' occupancy gives information consistent with the above and we do not analyze further.

\section{Summary and Discussion}
We use the pattern language to explore the SS model with a small lattice size of $L = 4\times 4$. By diagonalizing the Hamiltonian in an operator space, we obtain six kinds of spin-patterns. Then we employ spin-patterns to identify the ground state phases. Our results show that the ground state could be qualitatively divided into three regions, where region I at small $\kappa$ and region III at large $\kappa$ are identified as the N{\'e}el AFM phase and the DS one, respectively. The intermediate regime with maximal frustration has a feature of diagonal two-domain structure. At $\alpha_{c1} \approx 1.277$, the SS model evolves from the the N{\'e}el AFM phase to diagonal two-domain dominated phase. At $\alpha_{c2} \approx 1.5$, the model occurs a first-order phase transition from diagonal two-domain dominated phase to the DS one.

Increasing lattice size, the pattern will show more abundant features, such as more domain structures. However, the phenomena that the phase diagram evolves from the N{\'e}el AFM state to the intermediate state, then to the DS state with $\kappa$ increasing does not change, only the phase transition points are slightly shifted. This will be testified in the future study.

\section{Acknowledgments}
The work is partly supported by the National Key Research and Development Program of China (Grant No. 2022YFA1402704) and the programs for NSFC of China (Grant No. 11834005, Grant No. 12247101).



%

\end{document}